# GRIPP: an automated computer-aided design tool for generating 3D printable gradient index acoustic devices

Yangbo Xie,[1,a)] Steven A. Cummer,[1]

[1] *Department of Electrical and Computer Engineering, Duke University, Durham, North Carolina 27708, USA*

Gradient index (GRIN) acoustic devices have spatially inhomogeneous refractive index profile and allow flexible control of the propagation of acoustic waves. Previous GRIN acoustic lenses are mostly inherently two-dimensional designs that are difficult to be extended to all three dimensions. Besides, manually designing the spatially inhomogeneous structure is both time-consuming and error-prone. In this work, we proposed and numerically verified an automated computer-aided design tool: GRadient Index Pick-and-Place (GRIPP) algorithm, for generating three-dimensional GRIN acoustic wave controlling devices with scalable and 3D printable structures. The algorithm receives as inputs a spatial distribution of refractive index and a pre-defined library of gradient index unit cells, and outputs a 3D model of GRIN device that is ready to be 3D printed. The tool enables rapid design and realization of a large variety of 3D GRIN acoustic devices, which can be useful in areas such as speaker system design, airborne ultrasonic sensing, as well as therapeutic ultrasound.

## I. INTRODUCTION

GRIN lenses offer flexible wave controlling functions that are difficult for conventional homogeneous lenses to achieve [1]. However, the spatial inhomogeneity of acoustic properties of GRIN lenses makes them difficult to design and fabrication with conventional materials readily available from nature or chemical synthesis. The emergence of sonic crystals and acoustic metamaterials makes possible the realizations of many theoretical acoustic GRIN designs [2-5]. Nevertheless, two limitations of the previous design methods hinder their further applications. First, most of the previous designs are inherently two-dimensional and the extension to the third dimension is difficult. Secondly, the manual designing of spatially inhomogeneous structures are very time-consuming (especially for those designs that lack spatial symmetries) and often prone to human errors.

In this paper, we present a universal computer-aided design tool for designing three-dimensional gradient index acoustic devices to address the above-mentioned challenges. The CAD tool is termed as 'GRadient Index Pick and Place (GRIPP)' algorithm and it is designed to operate in LiveLink™ for MATLAB environment for the convenience of usage. Firstly, an analytic expression of the refractive index profile as well as a library of gradient index unit cells are fed into the algorithm; Secondly, the algorithm automatically discretizes the 3D space into a spatial grid with subwavelength grid cells; Thirdly, the algorithm loops over each grid cell and successively fill it with a unit cell that optimally matches the desired local refractive index. Eventually, the algorithm outputs a 3D model in the form of a STL file that can be directly sent to a 3D printer for fabrication.

## II. ALGORITHM DESCRIPTION

We describe in this section the workflow of the GRIPP algorithm. The algorithm receives two inputs from the user: an expression of the continuous spatial distribution of refractive index, and a library of gradient index unit cells covering the required range of refractive index. As shown in Fig. 1, first, a continuous function of the three-dimensional profile of the refractive index $n(x, y, z)$ and a library of gradient index unit cells are fed into the GRIPP. Second, the algorithm discretizes the continuous refractive index profile into spatial grids with a user-defined subwavelength grid cell size. Third, GRIPP scans through all the grid points, inquire the local refractive index at each grid point, and then search for the unit cell with the closest match of refractive index in the pre-defined library, and then 'pick-and-place' the selected unit cell from the library to the grid point. (The process is similar to the 'pick-and-place' process in the automated surface-mount electronic element assembly in the printed circuit board fabrication.) The code is implemented in MATLAB environment and LiveLink™ for MATLAB module in used to utilize COMSOL's built-in computer-aided design functionality for the 3D geometry generation and STL file exporting.

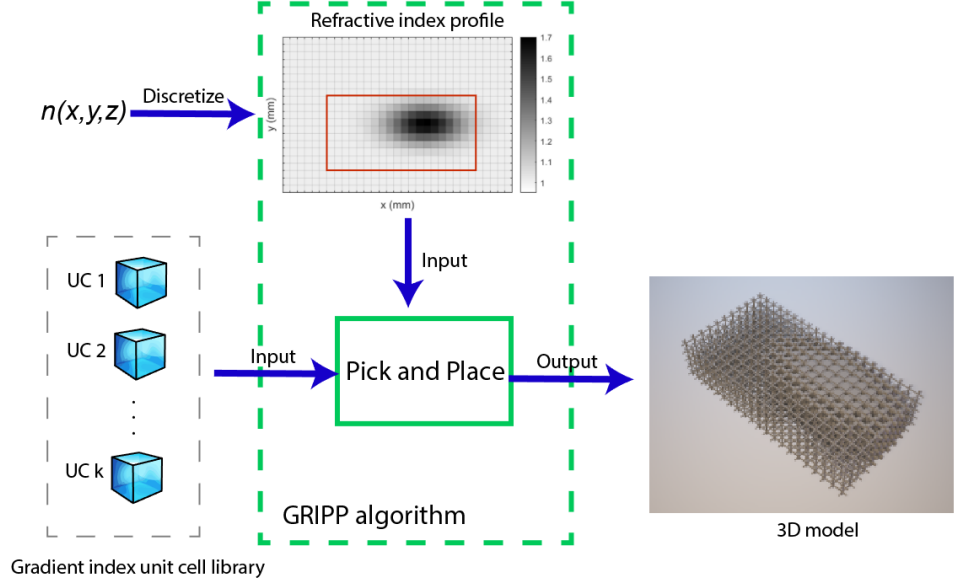

FIG. 1. The schematic of the workflow of the GRIPP algorithm.

## III. AN EXAMPLE: A 2D GRIN LENS

Below we will demonstrate the GRIPP algorithm with an example of GRIN lens. The refractive index profile of the lens can be expressed as

$$n(x, y, z) = 1 + 0.7 e^{-\frac{(x-x_c)^2+(y-y_c)^2}{r_0^2}} \quad (1)$$

where $x_c = 7.5mm$, $y_c = 2.5mm$ and $r_0 = 5mm$. We define the spatial grid cell size to be 2mm, or about a quarter of wavelength at the interested frequency of 40 kHz. The refractive index profile of this lens is shown in the central inset of Fig. 1.

The above refractive index profile requires the refractive index to have a range between 1 and 1.7. Many unit cell designs are possible to achieve this moderate range, here we use as an example a series of designs shaped as 3D-cross. As shown in Fig. 2, the 3D-cross unit cell has three orthogonal stubs. In principle, a range of anisotropic refractive index can be achieved with this design with different dimensions along different directions. Since our example of a 2D GRIN lens has only isotropic refractive index profile, we simplify the dimensions and let $a_x = a_y = a_z = a_0$. We also define $d_x = d_y = d_z = D$, where D is the size of the grid cell, so that each cell is interconnected with its adjacent cells to form a self-supporting lattice. The pre-defined library contains 12 unit cells that covers the range of refractive index between 1 and about 1.77, as shown in Table I.

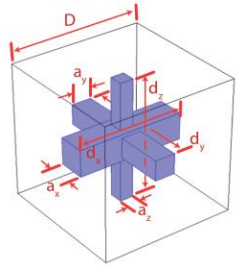

FIG. 2. The dimensions of an example 3D-cross gradient index unit cell.

TABLE I. Parameters for the employed library of the gradient index unit cell.

| Unit cell index | $a_0$ | Filling ratio | $n$ | $Z$ |
|---|---|---|---|---|
| 1 | 0.05 | 0.725% | 1.0024 | 1.0131 |
| 2 | 0.10 | 2.800% | 1.0102 | 1.0296 |
| 3 | 0.20 | 10.400% | 1.0456 | 1.2249 |
| 4 | 0.30 | 21.600% | 1.1133 | 1.5765 |
| 5 | 0.40 | 35.200% | 1.2087 | 2.2369 |
| 6 | 0.45 | 42.525% | 1.2661 | 2.7717 |
| 7 | 0.50 | 50.000% | 1.3320 | 3.5503 |
| 8 | 0.55 | 57.475% | 1.3973 | 4.6700 |
| 9 | 0.60 | 64.800% | 1.4740 | 6.4654 |
| 10 | 0.65 | 71.825% | 1.5644 | 9.5874 |
| 11 | 0.70 | 78.400% | 1.6607 | 15.370 |
| 12 | 0.75 | 84.375% | 1.7719 | 28.242 |

To verify the performance of the design generated by the GRIPP algorithm, we compared the simulated results between an ideal GRIN lens with continuous refractive index profile given by (1) and unity impedance, as well as that of a GRIN lens with a real structure consisting of interconnected 3D-cross cells. The control result of the ideal lens is shown in Fig. 3a, where the bended focus can be clearly identified. The result with the structured lens is shown in Fig. 3b (and Fig. 3c is a zoomed-in view of the geometry of the real structure). Excellent agreement is achieved between these two simulations. Minor differences between these two simulated results are likely caused by the non-unity impedance of the unit cells with 3D-cross structures. As shown in Table I, the impedance of the unit cells become more than 5 when the refractive index goes beyond 1.5. However, the gradient index design has the advantage of smooth transitioning of impedance difference, which essentially acts as an impedance matching network to reduce the undesired scattering caused by the impedance mismatch.

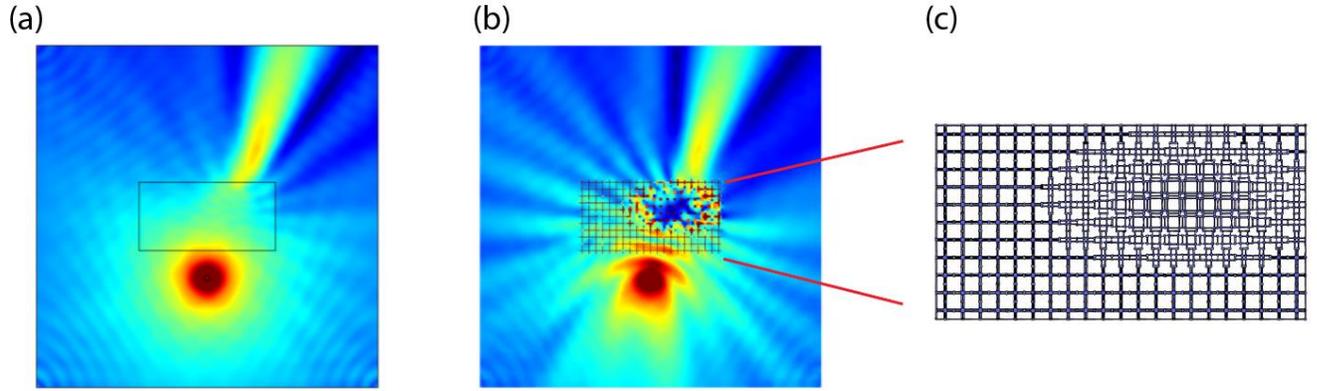

FIG. 3. The top-down view of the 3D simulation results of (a) an ideal GRIN lens (continous refractive index and unity impedance), (b) a GRIN lens with real 3D-cross structures (colormap is scaled to the same as (a)), and (c) is the zoomed-in top-down view of the geometery of the GRIN lens.

## IV. CONCLUSION

In conclusion, we presented here a computer-aided design tool known as GRIPP algorithm for generating 3D printable gradient index (GRIN) acoustic devices. An expression of the desired three-dimensional refractive index profile and a pre-defined library of unit cells with gradient refractive index are used as the inputs to the GRIPP algorithm, which then scan through the whole spatial grids, pick the unit cell with the best match and place it to the grid point. The final output from the algorithm is a 3D model of the structure that can be directly sent for fabrication.

With its versatility and convenience, GRIPP algorithm would be useful for the rapid design and realization of a large variety of three-dimensional GRIN acoustic devices. The algorithm may be extended to applications in electromagnetics to design 3D antennas and lenses.

## ACKNOWLEDGMENTS

Y. X. and S. A. C wishes to thank the support by the Multidisciplinary University Research Initiative grant from the Office of Naval Research (N00014-13-1-0631) and an Emerging Frontiers in Research and Innovation grant from the National Science Foundation (Grant No. 1641084).

## REFERENCES

1. N. Kundtz and D. R. Smith, "Extreme-angle broadband metamaterial lens," Nature Materials **9**(2), 129-132 (2010).
2. T. P. Martin, M. Nicholas, G. J. Orris, L. W. Cai, D. Torrent, and J. Sánchez-Dehesa, "Sonic gradient index lens for aqueous applications," Applied Physics Letters, **97**(11), 113503 (2010).
3. A. Climente, D. Torrent, and J. Sánchez-Dehesa, "Sound focusing by gradient index sonic lenses," Applied Physics Letters, **97**(10), 104103 (2010).

4. R. Q. Li, B. Liang, Y. Li, W. W. Kan, X. Y. Zou, and J. C. Cheng, "Broadband asymmetric acoustic transmission in a gradient-index structure," Applied Physics Letters, **101**(26), 263502 (2012).
5. Y. Ye, M. Ke, Y. Li, T. Wang, and Z. Liu, "Focusing of spoof surface-acoustic-waves by a gradient-index structure," Journal of Applied Physics, **114**(15), 154504 (2013).